# A Gate-All-Around Single-Channel In$_2$O$_3$ Nanoribbon FET with Near 20 mA/μm Drain Current

Zhuocheng Zhang[1,+], Zehao Lin[1,+], Pai-Ying Liao[1], Vahid Askarpour[3], Hongyi Dou[2], Zhongxia Shang[2], Adam Charnas[1], Mengwei Si[1], Sami Alajlouni[1], Jinhyun Noh[1], Ali Shakouri[1], Haiyan Wang[2], Mark Lundstrom[1], Jesse Maassen[3], and Peide D. Ye[1,*]

[1]School of Electrical and Computer Engineering, [2]School of Materials Science and Engineering, Purdue University, West Lafayette, U.S.A.
[3]Department of Physics and Atmospheric Science, Dalhousie University, Halifax, NS, Canada
[+]These authors contributed equally to this work *Email: yep@purdue.edu

## Abstract
In this work, we demonstrate atomic-layer-deposited (ALD) single-channel indium oxide (In$_2$O$_3$) gate-all-around (GAA) nanoribbon FETs in a back-end-of-line (BEOL) compatible process. A maximum on-state current ($I_{ON}$) of 19.3 mA/μm (near 20 mA/μm) is achieved in an In$_2$O$_3$ GAA nanoribbon FET with a channel thickness ($T_{IO}$) of 3.1 nm, channel length ($L_{ch}$) of 40 nm, channel width ($W_{ch}$) of 30 nm and dielectric HfO$_2$ of 5 nm. The record high drain current obtained from an In$_2$O$_3$ FET is about one order of magnitude higher than any conventional single-channel semiconductor FETs. This extraordinary drain current and its related on-state performance demonstrate ALD In$_2$O$_3$ is a promising oxide semiconductor channel with great opportunities in BEOL compatible monolithic 3D integration.

## Introduction

Amorphous oxide semiconductors are widely investigated and applied as thin-film transistor channels for display applications [1]. Very recently, they are explored for BEOL-compatible transistor channels for monolithic 3D integration [2-5]. Among them, ALD In$_2$O$_3$ shows excellent transport properties with mobility beyond 100 cm$^2$/V·s [6] and its FETs exhibit remarkable performance including large on-state current ($I_{ON}$) over 2-3 mA/μm in both depletion and enhancement-mode operation, on/off ratio up to 10$^{17}$, subthreshold swing (SS) as low as 63 mV/dec, and high stability in H$_2$ environment [6-10]. Meanwhile, the ALD technique offers wafer-scale film synthesis, accurate thickness control, smooth surface, high conformity and uniformity on 3D structures and a low thermal budget process below 300 °C [6-8,11-13].

In this work, for the first time, scaled single-channel In$_2$O$_3$ nanoribbon FETs with GAA structure are fabricated under BEOL conditions. $I_{ON}$ is enhanced to 4.3 mA/μm for a channel length ($L_{ch}$) of 40 nm compared to previous bottom or top-gated ALD In$_2$O$_3$ FETs. To resolve the self-heating issue due to the ultrahigh current in the ultrathin channel, the channel width ($W_{ch}$) is further scaled from 1 μm to 30 nm. The self-heating effect is mitigated significantly after nanoribbon scaling as confirmed by a high-resolution thermo-reflectance (TR) imaging experiment and thermal simulation. A maximum $I_{ON}$ of 19.3 mA/μm is observed in the 30 nm wide nanoribbon FET by pulsed I-V measurements, which further lowers the self-heating effect. This extraordinary drain current density is the highest ever obtained from a single-channel transistor.

## Experiments

Fig. 1 shows the device schematic of a single-channel In$_2$O$_3$ GAA FET. A fabrication process flow is presented in Fig. 2. The device fabrication started with ALD growth of 10 nm Al$_2$O$_3$ at 175 °C to obtain a smooth surface and 40 nm Ni deposition by e-beam evaporation as the bottom gate metal after substrate cleaning. Next, 5 nm HfO$_2$ bottom dielectric was grown by ALD at 200 °C and 3.1 nm In$_2$O$_3$ was grown by ALD at 225 °C with (CH$_3$)$_3$In (TMIn) and H$_2$O as In and O precursors. Channel isolation was done by dry etching (BCl$_3$: 15 sccm; Ar: 60 sccm; pressure: 0.6 Pa) and Ni was deposited to serve as the source/drain electrodes. A top dielectric of 5 nm HfO$_2$ was then grown by ALD at 200 °C to wrap the whole In$_2$O$_3$ channel followed by 5 min O$_2$ annealing at 250 °C. Finally, the gate metal was surrounded by e-beam evaporation of 40 nm Ni with dry etching first to connect the top and bottom gates. Fig. 3 shows the EDS cross-section image of an In$_2$O$_3$ GAA FET. Single layer In$_2$O$_3$ with channel thickness ($T_{IO}$) of 3.1 nm is surrounded by 5 nm HfO$_2$ and 40 nm Ni gate stack.

## Results and Discussion

Fig. 4 and Fig. 5 present the transfer and output characteristics of a typical In$_2$O$_3$ GAA FET with $L_{ch}$ of 1 μm and $W_{ch}$ of 1 μm. Well-behaved switching performance with on/off ratio over 10$^4$ and a clear drain current saturation at large $V_{DS}$ are observed. Fig. 6 and Fig. 7 show the transfer and output characteristics of a short channel In$_2$O$_3$ GAA FET with $L_{ch}$ of 40 nm and $W_{ch}$ of 1 μm. A maximum $I_{ON}$ of 4.3 mA/μm is achieved at $V_{DS}$ of 1 V. Further increasing $V_{DS}$ results in the device becoming unstable due to ultrahigh current density induced Joule heating in the nanometer-thin In$_2$O$_3$ channel. The threshold voltage ($V_T$) and transconductance ($g_m$) scaling metrics of the In$_2$O$_3$ FETs are shown in Fig. 8. $V_T$ is near constant and maximum $g_m$ exceeds 10$^3$ μS/μm.

Channel width scaling of the ribbon is also investigated. Fig. 9 illustrates a false-color top-view SEM image of a fabricated GAA In$_2$O$_3$ nanoribbon FET with $W_{ch}$ of 50 nm. Electrical measurements are performed and the extracted $I_{ON}$ versus $W_{ch}$ of different devices from 1 μm down to 30 nm is plotted in Fig. 10 with subthreshold swing (SS) of 100-120 mV/dec. It is surprising that $I_{ON}$ is approximately three times larger as the channel narrows. Two factors might account for such current enhancement. First, the electric field induces a higher carrier density at the channel edge due to the GAA geometry and the edge gradually takes a larger portion of the conductance in the channel width scaling. Etched In$_2$O$_3$ ribbon edges lead to a charge neutrality level (CNL) located even deeper inside the conduction band, thereby enhancing edge conduction. Second, better heat dissipation is realized in a narrower ribbon. Fig. 11 and Fig. 12 present the transfer and output characteristics of an In$_2$O$_3$ FET with $L_{ch}$ of 40 nm and $W_{ch}$ of 30 nm. A maximum $I_{ON}$ of 11.4 mA/μm is achieved at a $V_{DS}$ of 1 V.

A high-resolution TR measurement is conducted to quantitatively investigate the self-heating effect [12]. The device is illuminated by LED pulses with a synchronized charge coupled device camera to capture the surface reflectance, which is then transformed into a temperature scale. Fig. 13 shows the experimental and simulated results of the temperature increase ΔT across the channel of an In$_2$O$_3$ GAA FET with $L_{ch}$ of 400 nm, $W_{ch}$ of 1 μm, $V_{DS}$ at 1.4 V and power density of 1.94 kW/mm$^2$. Thermal simulation is carried out by the finite-element method using COMSOL. A good agreement is found between experiment and simulation with a maximum ΔT of 23.6 K. To compare with narrow devices, Fig. 14 presents the experimental and simulated results of another In$_2$O$_3$ GAA nanoribbon FET with $W_{ch}$ of 200 nm and the same $V_{DS}$ bias. It can be seen that the maximum ΔT reduces to 5.1 K due to better heat dissipation in the smaller nanoribbon. Fig. 15 shows the simulated 2D ΔT results of the smallest In$_2$O$_3$ GAA nanoribbon FET, with $L_{ch}$ of 40 nm and $W_{ch}$ of 30 nm, in which a negligible ΔT of 1.1 K is achieved. The excellent transport properties of In$_2$O$_3$ remain unchanged in scaled nanoribbon devices.

To further probe the potential of the device performance, a pulsed I-V measurement is implemented. A data averaging time of 500 ns, $V_{GS}$ and $V_{DS}$ pulse width of 1 μs and pulse period of 100 ms are set to minimize the self-heating effect and improve device reliability. Fig. 16 and Fig. 17 present the pulsed transfer and output characteristics of an In$_2$O$_3$ GAA nanoribbon FET with $L_{ch}$ of 40 nm and $W_{ch}$ of 30 nm. A record high $I_{ON}$ of 19.3 mA/μm is achieved at $V_{DS}$ of 1.7 V, demonstrating the remarkable current carrying capacity of ALD In$_2$O$_3$. The off-state current in the transfer curve is limited by the resolution of the pulsed I-V method while a real on/off ratio over 10$^6$ is shown in Fig. 11. DFT simulations [13] of a 3.0 nm thick slab of In$_2$O$_3$ were performed to study its electronic properties. Fig. 18 shows the band structure of the conduction states. Fig. 19 presents the ballistic current density and average electronic velocity versus carrier concentration, indicating that an $I_{ON}$ of 20 mA/μm with high velocity >3×10$^7$ cm/s can be achieved with an electron density of 4×10$^{13}$ cm$^{-2}$ [14].

## Conclusion

In summary, single-channel In$_2$O$_3$ GAA nanoribbon FETs are demonstrated in a BEOL compatible process. A record high $I_{ON}$ of 19.3 mA/μm is achieved through channel width scaling and improved heat dissipation in small devices. This work demonstrates the great potential of ALD In$_2$O$_3$ as a novel channel material in monolithic 3D integration.

The work is supported by SRC nCore IMPACT, DARPA/SRC JUMP ASCENT, AFOSR, NSERC and Compute Canada.

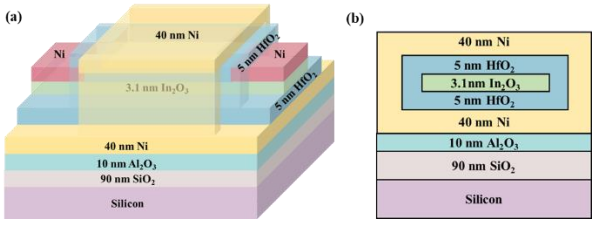

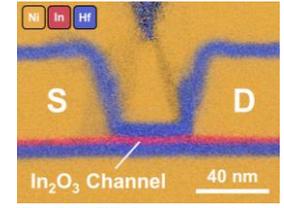

Fig. 1. Device schematic of an $In_2O_3$ GAA nanoribbon FET with $T_{IO}$ of 3.1 nm and dielectric of 5 nm $HfO_2$ in (a) 3D model (b) cross-section view in S/D direction.

Fig. 2. Fabrication process flow of the $In_2O_3$ GAA nanoribbon FETs.

Fig. 3. EDS cross-section image of an $In_2O_3$ GAA FET with $L_{ch}$ of 40 nm.

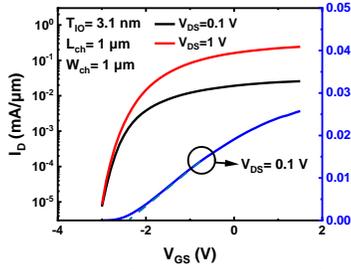
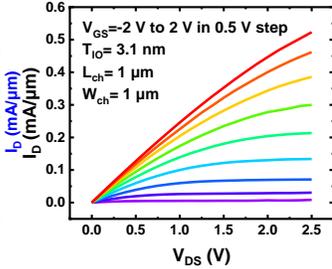
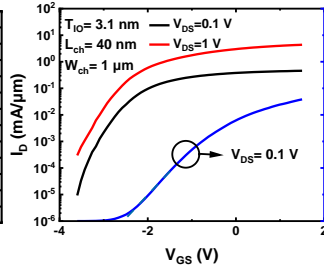
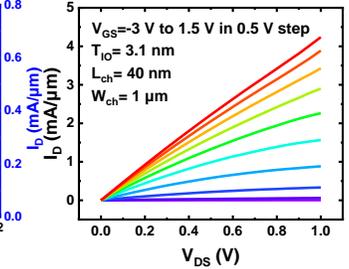

Fig. 4. Transfer characteristics of a typical $In_2O_3$ GAA nanoribbon FET with $L_{ch}$ of 1 μm and $W_{ch}$ of 1 μm.

Fig. 5. Output characteristics of a typical $In_2O_3$ GAA nanoribbon FET with $L_{ch}$ of 1 μm and $W_{ch}$ of 1 μm, showing saturation at large $V_{DS}$.

Fig. 6. Transfer characteristics of a typical $In_2O_3$ GAA FET with $L_{ch}$ of 40 nm and $W_{ch}$ of 1 μm.

Fig. 7. Output characteristics of a typical $In_2O_3$ GAA FET with $L_{ch}$ of 40 nm and $W_{ch}$ of 1 μm, showing maximum $I_{ON}$ of 4.3 mA/μm at $V_{DS}= 1$ V.

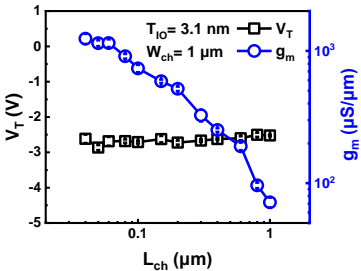
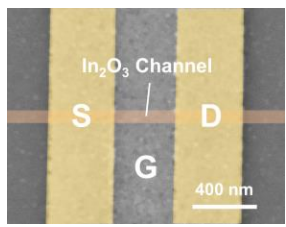
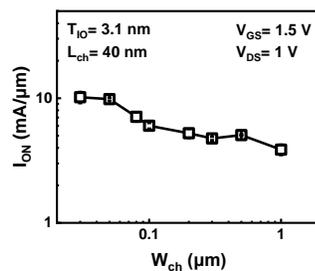
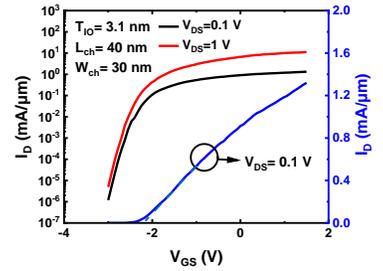

Fig. 8. $V_T$ and $g_m$ scaling metrics of $In_2O_3$ GAA FETs with $L_{ch}$ from 40 nm to 1 μm and $W_{ch}$ of 1 μm. $g_m$ is extracted at $V_{DS}=1$ V.

Fig. 9. False-color top-view SEM image of an $In_2O_3$ GAA nanoribbon FET with $W_{ch}$ of 50 nm.

Fig. 10. $I_{ON}$ of the $In_2O_3$ GAA nanoribbon FETs under channel width scaling from 1 μm down to 30 nm. Each $W_{ch}$ data is from 5-10 devices.

Fig. 11. Transfer characteristics of a typical $In_2O_3$ GAA nanoribbon FET with $L_{ch}$ of 40 nm and $W_{ch}$ of 30 nm.

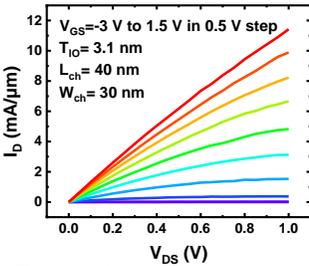
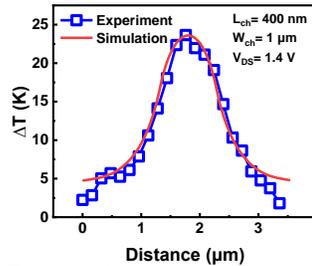
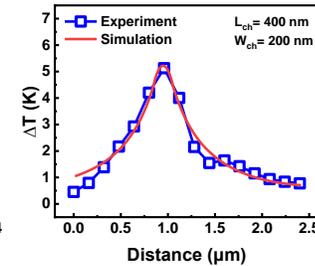
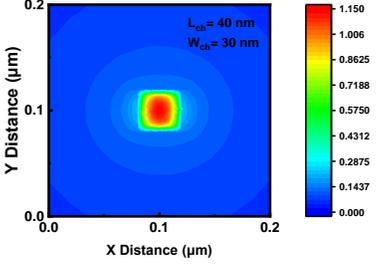

Fig. 12. Output characteristics of an $In_2O_3$ GAA nanoribbon FET with $W_{ch}$ of 30 nm, showing maximum $I_{ON}$ of 11.4 mA/μm at $V_{DS}= 1$ V.

Fig. 13. Cross-sections of the experimental and simulated results of temperature increase ΔT of an $In_2O_3$ GAA nanoribbon FET with $W_{ch}$ of 1 μm.

Fig. 14. The same TR measurements and simulations of a device with $W_{ch}$ of 200 nm. Space resolution is limited to hundreds of nm for TR measurement.

Fig. 15. Simulation result of an $In_2O_3$ GAA nanoribbon FET with $L_{ch}$ of 40 nm and $W_{ch}$ of 30 nm, showing little temperature increase.

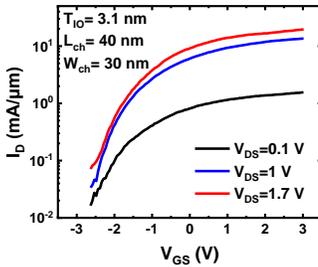
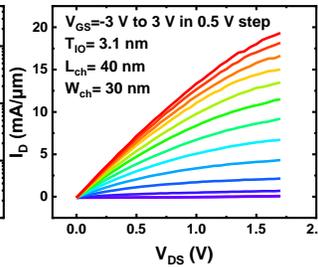
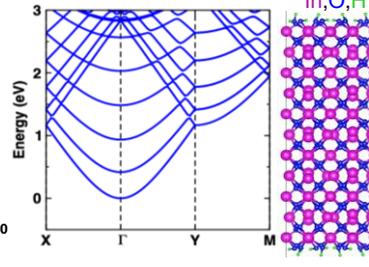
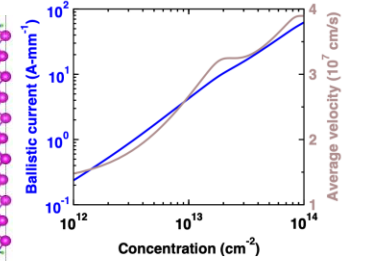

Fig. 16. Pulsed I-V results of transfer characteristics of an $In_2O_3$ GAA FET with $L_{ch}$ of 40 nm and $W_{ch}$ of 30 nm.

Fig. 17. Pulsed I-V results of output characteristics of an $In_2O_3$ GAA FET, showing maximum $I_{ON}$ of 19.3 mA/μm.

Fig. 18. DFT band structure of 3.0 nm thick $In_2O_3$ slab, with the lowest conduction band having an effective mass of 0.19 $m_0$.

Fig. 19. Ballistic current density and average electron velocity versus carrier density, from DFT simulation of $In_2O_3$ slab.